# Correlations between spin accumulation and degree of time-inverse breaking for electron gas in solid

*V.Zayets**

*Spintronic Research Center, National Institute of Advanced Industrial Science and Technology (AIST), Umezono 1-1-1, Tsukuba, Ibaraki, Japan. E-mail: v.zayets@aist.go.jp*

*It is shown that the electron spin may not be conserved after a spin-independent scattering. This fact strongly limits the validity of the classical model of spin-up/spin-down bands, which has been used for description of magnetic properties of conduction electrons. It is shown that it is possible to divide all conduction electrons into two group distinguished by their symmetry for time reversal. The number of electrons in each group is conserved after a spin-independent scattering. This makes it convenient to use these groups for describing of the magnetic properties of conduction electrons. The energy distribution of spins, the Pauli paramagnetism and the spin distribution in the ferromagnetic metals are described within the presented model. The effects of spin torque and spin-torque current are described. The origin of spin-transfer torque is explained within presented model.*

## 1. Introduction

Spintronics is a new type of electronics that exploits the spin degree of freedom of an electron in addition to its charge. There are many expectations that in the near future, spintronics devices will be competitive with modern Si electronics devices. It is expected that spintronics devices will be faster, compacter and more energy-saving.

In the last decade there have been significant advances in the field of spintronics. New effects, new functions and new devices have been explored. Spin polarized current was efficiently injected from a ferromagnetic metal into a non-magnetic metal and a semiconductor[1-5], the method of electrical detection of spin current was developed[6,7], the Spin Hall effect[8] and the inverse Spin Hall effect[9] were experimentally measured, the operation of a spin transistor[10] and tunnel spin transistor[11-13] were experimentally demonstrated, the tunnel junction with magneto-resistance over 100 % was developed[14,15].

However, it is still difficult for spintronics devices to compete with modern Si devices. Optimization of spintronics devices is important in order to achieve the performance required for commercialization. The modeling of spin transport and the understanding of spin properties of conduction electrons in metals and semiconductors are a key for such optimization.

The classical model used for the description of the magnetic properties of conduction electrons in metals is the model of spin-up/spin-down bands. In this model it is assumed that all conduction electrons in a solid occupy two bands, which are named spin-up and spin-down bands. The electrons of only one spin direction occupy each band. Based on the fact that the probability of a spin-flip scattering is low, it was assumed that the electrons are spending a long time in each band and the exchange of electrons between the bands is rare. Using this key assumption the classical model describes the transport and magnetic properties of conduction electrons independently for each spin band. It includes only a weak interaction between the electrons of the spin-up and spin-down bands. For example, in the classical Pauli's description of paramagnetism in non-magnetic metals[16], it is assumed that electrons in one band have spin parallel and electrons in another band have spin anti parallel to the direction of the magnetic field. Without a magnetic field there is an equal number of electrons in each band and the total magnetic moment of the electron gas is zero. In a magnetic field the electrons with spin parallel to the magnetic field (spin-up) have lower energy compared to the case with no magnetic field, and electrons with spin anti parallel to the magnetic field have higher energy. In equilibrium the Fermi energy is the same for both bands. This means that in a magnetic field some electrons had spin-flipped from the spin-down band to the spin-up band and so the number of electrons in the spin-up band became larger than in the spin-down band. Because of this difference, the magnetic moment of the electron gas became non-zero.

It has been assumed that the spin-flip from one spin band to another takes a relatively long time. This makes it possible to define the separate Fermi energies for spin-up and spin-down bands in the classical model. The energy distribution in each spin band is described by the Fermi-Dirac distribution with different Fermi energies. The Fermi energies of the spin-up and spin-down bands relax into a single Fermi energy within a certain time, which is called the spin relaxation time.

Because of the weak interaction between the electrons of the different spin bands, the electron transport in each band is considered to be independent of the electrons of the other spin band. The classical model makes possible the description of the transport of each spin band by two separate Ohms laws with different conductivities[17,18] for spin-up and spin-down electrons. For example, it was assumed that the density of states in a ferromagnetic metal is different for spin-up and spin-down bands near its Fermi level. Therefore, the number of electrons participating in the transport in the metal is different for the spin-up and spin-down bands. This leads to a difference in the conductivities of spin-up and spin-down electrons. Another example is electron transport in a non-degenerated semiconductor in which spin is accumulated. The number of electrons in the conduction band of a semiconductor depends on the position of the Fermi energy relative to the conduction band minimum. In the case of spin accumulation, according to the classical model of spin-up/spin-down bands, the Fermi energies for spin-up and spin-down electrons should be different. This leads to a difference in the number of spin-up and spin-down electrons and correspondingly to different conductivities for spin-up and spin-down electrons in the semiconductor[18].

In Chapter 2 it will be shown that even in the case of a spin-independent scattering the spin direction of a conduction electron is not conserved and its spin may rotate. This feature is due to the quantum of nature of spin. The scatterings, after which spin direction is not conserved, are rather frequent. If at some moment in time the electrons only have two opposite directions of spin, within a very short time the spin-independent scatterings will mix up all electrons and there will be electrons with all possible spin directions. This fact questions the validity of the classical model of spin-up/spin-down bands.

Even though the spin direction is not conserved after spin-independent scatterings, the time-inverse symmetry/asymmetry is conserved. In Chapter 3 it will be shown that it is possible to divide all conduction electrons into two groups, which are defined as the time-inverse symmetrical (TIS) assembly and the time-inverse asymmetrical (TIA) assembly. In the TIS assembly the spin can have any possible direction with equal probability. In the TIA assembly all electron spins are in one direction. Even though the exchange of electrons between assemblies is frequent, the spin-independent scatterings do not change the number of electrons in each assembly.

In Chapter 4, the energy distribution of electrons in TIA and TIS assemblies is described. Since in the classical model of spin-up/spin-down bands only a weak interaction between electrons of different spin bands is assumed, the model allows the energy distribution of the total number of electrons to be different from the Fermi-Dirac statistic. It only requires that the energy distribution of the electrons in each band follows the Fermi-Dirac distribution. The proposed model of TIS/TIA assemblies gives a different energy distribution than the classical model. Since spin-independent scatterings intermix all electrons, including the electrons of different assemblies, the distribution of the total number of electrons should be described by a single Fermi-Dirac distribution even in the case of a spin accumulation. Using this fact and the properties of spin-scatterings the energy distribution of electrons in each assembly is described in Chapter 4.

The presented model of TIA and TIS assemblies describes the magnetic properties of an electron gas in terms of a conversion of electrons between the assemblies. For example, in equilibrium all conduction electrons of a non-magnetic metal are in the TIS assembly. When there is a spin accumulation in the metal, it means that there are some electrons in the TIA assembly. The relaxation of the spin accumulation corresponds to a conversion of electrons from the TIA assembly into the TIS assembly.

In Chapter 5, it was shown that in the presence of the magnetic field the electrons of the TIS assembly are converted into TIA assembly. For example, if without magnetic field all conduction electrons in a non-magnetic metal are only in the TIS assembly, under a magnetic field there are electrons in both TIA and TIS assemblies. The number of electrons in each assembly is determined by the condition, that the conversion rate of electrons from TIS assembly into TIA assembly induced by the magnetic field should be balanced by the reverse conversion due to a finite spin life time. Since the magnetization of the TIA assembly is non-zero, the applied magnetic field induces a magnetization in the electron gas. This effect is called the Pauli paramagnetism of electron gas and it is described in Chapter 6. The Pauli paramagnetism occurs because a magnetic field induces a spin accumulation in a non-magnetic metal and the spin polarization of the non-magnetic metal becomes non-zero.

In Chapter 7 the properties of conduction electrons in a ferromagnetic metal are described. In equilibrium the conduction electrons in a ferromagnetic metal are in both TIA and TIS assemblies. The number of electrons in each assembly is determined by the condition, that the conversion rate of electrons from the TIS into the TIA assembly, induced by the exchange interaction between local d-electrons and the sp-electrons, is balanced by the reverse conversion due to a finite spin life time in the ferromagnetic metal.

In a metal one TIS and one TIA assembly can coexist for a relatively long time, while two TIA assemblies will quickly combine. It is possible that at some moment in time in a metal there are two or more TIA assemblies. However, within a short time the assemblies combine into one TIA assembly and some electrons are converted into TIS assembly. The temporal evolution of the interaction of two TIA assemblies is described in Chapter 8. In this chapter the interesting case of the interaction of two TIA assemblies is also described. In this case in a metal there is a substantial number of electrons in one TIA assembly. A small amount of electrons of the other TIA assembly are injected into the metal or converted from the TIS assembly at some rate. As a result the spin direction of the TIA assembly rotates towards the spin direction of the injected electrons. The torque acting on the electrons of the TIA assembly is named the spin torque.

In Chapter 9 the case, when the spin direction of the TIA assembly is different at different spacial points in the metal, is studied. In this case the electrons of different spin directions diffuse from point to point. This causes a spin torque. Such a flow of electrons is named the spin-torque current. The spin-torque current is trying to align the spin direction of the TIA assembly, so that it would be the same over whole sample. In contrast to the spin current, which is related to the diffusion of the spin accumulation, the spin-torque current is related to the diffusion of the direction of the spin accumulation.

In Chapter 10 the physical origin of the spin-transfer torque is described. The spin transfer torque is the torque acting on the magnetization of a ferromagnetic electrode of a magnetic tunnel junction (MTJ), when an electrical current flows through MTJ. The optimization of the spin transfer torque is an important task for a variety of practical applications. It was found that an essential feature of the spin-transfer torque is the non-zero angle between the spin directions of conduction electrons of the TIA assembly and the local d-electrons. The spin direction of the TIA assembly rotates out from the spin direction of the d-electrons, because of the spin-torque current flowing between the ferromagnetic electrodes.

## 2. Spin rotation after a spin independent scattering.

By the spin-independent scattering we define a scattering, which does not depend on the electron spin. That means there is no spin operator in the Hamiltonian describing a scattering event. In the following we will show that even in this case the electron spin may not be conserved during the scattering because of the quantum nature of spin.

A electron is a fermion with spin 1/2. Its quantum states is described by two wavefunctions corresponded to opposite directions of spin. An electron quantum state can either be filled by no electrons or one electron or two electrons of opposite spins. These three states we define as "full", "empty" and "spin" states (See Fig.1(a)). In the case of a "spin" state, one of the two states is occupied by an electron. The emptiness of the neighbor state allows a rotation of spin. In the "spin" state the electron has a spin direction and electron wave function that can be describe as

$$\Psi = a \cdot \Psi_\uparrow + b \cdot \Psi_\downarrow \qquad (1)$$

where $\Psi_\uparrow$ and $\Psi_\downarrow$ are the electron wave functions in the cases when spin is directed along and opposite to the direction of the z-axis, respectively.

The spinor $\hat{S}$ describes the coefficients in Eq.(1) as

$$\hat{S} = \begin{pmatrix} a \\ b \end{pmatrix} \qquad (2)$$

A change of the spin direction can be described by a corresponding change of the coefficients a and b in the spinor (2).
There are two electron of opposite spin in a "full" state. Therefore, the total spin of "full" state is zero. As an object with spin zero, the "full" state does not have any spin direction. Except for the fact that the spin directions of the two electrons in the "full" state are opposite, there is no other information about the spin direction. The spin of the electrons in "full" state may be considered as directed in any direction with equal probability (See Fig. 1b). For example, the three states shown in Fig. 1(b) are equivalent, undistinguishable and all of them equally represents one "full" state. Because of this quantum-mechanical property of a "full" state, the spin direction of conduction electrons is not conserved during a spin-independent scattering.

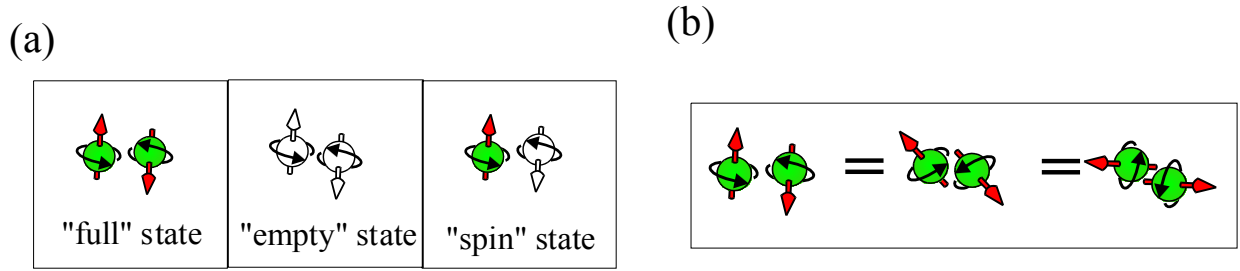

*Fig.1 (a) Three occupation possibilities of a quantum state: a "full" state is a state occupied by two electrons; an ''empty" state is a state, which is not occupied by any electrons; a "spin" state is a state, in which only one of the two states is occupied. (b) three representations of a "full" state which are undistinguishable. This is because spin of the "full" state is zero and the "full" state does not have a spin direction.*

For example, Figure 2 shows two consecutive scatterings of two spin states of opposite spin direction. Before scatterings the spin directions of the "spin" states are up and down. During the first spin-independent scattering, the electron from one "spin" state is scattered into an empty state of other "spin" state. The result of the first scattering is one "full" state and one "empty" state. The first scatterings converted two spin states, which have spin directions, into two states, which do not have any spin direction.

During the second scattering, one electron from a "full" state is scattered back into an "empty" state. The result of the second scattering is two "spin" states. After the second scattering the spin directions of the "spin" states are left and right. The spin directions are opposite, but there are different from the spin directions of the "spin" states before these two scatterings. Therefore, after two consecutive spin-independent scatterings the spin direction randomly changes.

The reason why spin is not conserved can be understood as follows. In the "full" state there are two electrons of opposite spin, but there is no information inside of the "full" state about the spin direction of electrons prior to being scattered into the "full" state. Also, when an electron is scattered out of a "full" state, it can have any spin direction with an equal probability. There is no correlation between the spin directions of electrons before and after scattering. An electron looses the information about its spin direction when it stays inside of "full" state.

The probability of the first scattering event of Fig.2 is not high, because it requires that two spin states of opposite spin are close to each other in space. Scatterings between two "spin" states, which have some angle $\phi$ between their spin directions, are more probable. This spin-independent scattering can result in either two "spin" states or a "full" plus an "empty" state. When spin directions of the initial "spin" states are parallel, the result of the scattering will only be two "spin" states. When spin directions of the initial "spin" states are anti parallel, the result of the scattering will only be "full" +"empty" states. For the other angles there is a probability that the scattering result is either two "spin" states or a "full" +"empty" state. In the following this probability is calculated.

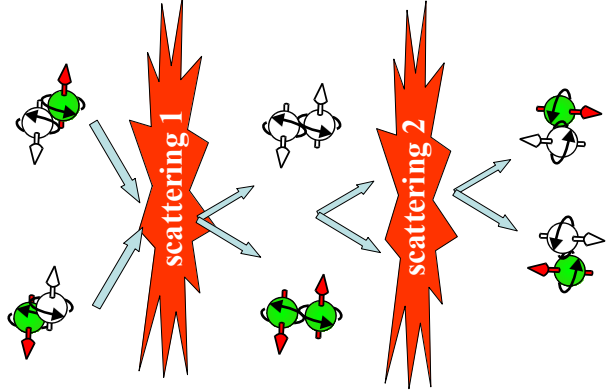

*Fig.2. Two consecutive scattering events, which result in the rotation of a spin direction*

Let us consider a scattering of two "spin" states. The spin directions of the "spin" states are in the xz-plane and have an angle $-\phi/2$ and $\phi/2$ with respect to the x-axis (See Fig.3). These "spin" states are described by spinors

$$\hat{S}_1 = \frac{1}{\sqrt{2}}\begin{pmatrix} 1 \\ e^{-i\phi/2} \end{pmatrix} \quad \hat{S}_2 = \frac{1}{\sqrt{2}}\begin{pmatrix} 1 \\ e^{i\phi/2} \end{pmatrix} \tag{3}$$

Two spin states can be considered as a closed system, the wavefunction of which is described by a product of the spinors S1 and S2. A spin-independent scattering can be considered as a perturbation inside the closed system and it does not change the overall wavefunction of whole closed system. Therefore, even after the scattering the two electrons are described by the same overall wavefunction as before the scattering. Even though wavefunction of each electron may change, the overall wavefunction of two electrons does not change. It is described by a product of the spinor S1 and S2.

The spin of a "full" and an "empty" state is zero. Therefore, the wavefunction of "full" or "empty" states is a scalar with respect to an axis rotation. There is only one possible scalar from a binary product of spinors S1 and S2. It is called the scalar product of spinors S1 and S2 (see Ref.19). Therefore, the wavefunction of "full" + "empty" states is the scalar product of spinors S1 and S2 and the probability that the scattered electrons will be in "full" + "empty" states is calculated as

$$p_{full+empty} = \left| S_1^i \cdot S_{2i} \right|^2 = \left| \frac{1 \cdot e^{i\phi/2} - 1 \cdot e^{-i\phi/2}}{2} \right|^2 = \sin^2(\phi/2) \tag{4}$$

Since the result of a scattering of two spin states can be either "full" +"empty" states or two "spin" states, from Eq. (4) the probability of scattered electrons to be in "spin" states is given as

$$P_{spin+spin} = 1 - p_{full+empty} = \cos^2(\phi/2) \tag{5}$$

It should be noted that in contrast to the above-discussed scattering of two spin states (Fig.2), there are two more spin-independent scatterings in which the spin direction is conserved. The first such scattering is the scattering in which an electron is scattered from a "spin" state into an "empty" state. The second scattering is the scattering of one electron from a "full" state into a "spin" state.

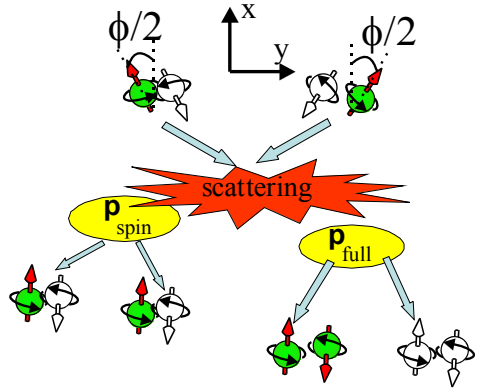

*Fig.3. Scattering of two "spin" states, which have angle $\phi$ between their spin directions. The result is either two "spin" states with parallel spins or a "full" state and an "empty" state.*

The scattering event shown in Fig.2, in which the spin direction is not conserved, has a high probability. For example, if at one moment all electrons have only spin-up or spin-down spin directions, within a short time the scattering event of Fig.2 will mix up all spins and spins will point into any arbitrary direction. Therefore, it is incorrect to describe electron gas as a mixture of electrons of only two opposite spin directions. In the electron gas it should be always electrons with spin directed in any direction. That fact significantly limits the validity of the classical model of spin-up/spin-down bands for the conduction electrons in a solid.

It should be noticed that the spin rotation mechanism shown in Figure (2) has a quantum-mechanical nature. It exists even in materials in which there are no intrinsic magnetic fields, no exchange interactions and no spin-orbital interactions.

## 3. TIS and TIA assemblies.

Even though the spin direction of one electron is not conserved after the scattering event shown in Fig.2, the total spin of all conduction electrons is conserved. Due to this fact it is possible to divide all conduction electrons into two groups. In the first group there are "spin" states with the same spin direction. The total spin of the first group is non-zero. In the second group there are "spin" states, "full" states and "empty" states. The total spin of the electrons of the second group is zero. As was shown in Appendix 1, even though there is a frequent exchange of electrons between the groups, the number of electrons in each group does not change due to spin-independent scatterings. This division of conduction electrons into two groups significantly simplifies the analysis of magnetic properties and spin-dependent transport of an electron gas.

It should be noticed that time reversal does not change the wavefunctions of "full" and "empty" states, because their spin is zero. In contrast, the "spin" states change their spin direction with time reversal. The time-inverse symmetry is an important property of the above-mentioned groups of electrons. We name these two groups as the time-inverse symmetrical (TIS) assembly and time-inverse asymmetrical (TIA) assembly. The TIS assembly contains the "full" and "empty" states, because they are time-inverse symmetrical. Even though "spin" states are time-inverse asymmetrical, the TIS assembly should contain them, because they are results of scatterings of an electron from a "full" state into an "empty" state (See the scattering event 2 in Fig.2). As long as there are "full" and "empty" states in an assembly, the assembly should contain "spin" states as well. In order for the TIS assembly to be time-inverse symmetrical, it should contain an equal number of "spin" states for all spin directions. The TIA assembly can not contain "full" and "empty" states, otherwise the scattering of Fig.2 will result in some "spin" states with different spin directions. Therefore, the TIA assembly contains only "spin" states with the same spin direction.

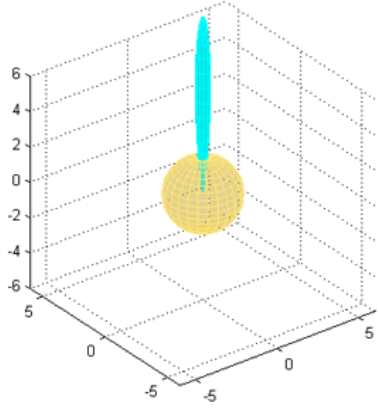

*Fig.4. The distribution of spin directions of "spin" states in the TIS (yellow) and in the TIA (blue) assembly. The length of a vector from the origin to the surface corresponds to the number of electrons, the spin direction of which is the same as the direction of the vector. All spins in the TIA assembly are directed in one direction. Spins in the TIS assembly are directed in all directions with equal probability.*

Figure 4 shows an example of the distribution of spin directions of "spin" states in the TIS and TIA assemblies. The angle distribution of the spin directions in the TIS assembly is a sphere. This is because of the fact that the TIS assembly should contain an equal amount of "spin" states with all spin directions. The distribution of the spin directions in the TIA assembly is like a delta-function with all spins in almost the same direction. There is a slight broadening because of spin relaxation.

It should be noticed that the division of conduction electrons into the TIA and TIS assemblies is only the possible division of electrons into groups, which are not mixed up by the spin-independent scatterings.

If there is a distribution of "spin" states with some arbitrary spin directions, within a short time the electrons are scattered and redistributed into one TIA assembly and one TIS assembly. A distribution with some arbitrary spin directions can be created, for example, by an external source. It is possible to calculate how many electrons will be in the final TIA and TIS assemblies. Let us assume that at some moment in time the angular distribution of spin directions is $p_{spin}(\theta)$, where $p_{spin}(\theta)$ is the probability for a "spin" state to have an angle between their spin direction and the z-axis in an interval $(\theta, \theta + d\theta)$. After the scatterings have redistributed the states into one TIA assembly and one TIS assembly, the probability of "spin" states to be in the TIA assembly can be calculated as (See Eq.(A1.7) in Appendix 1)

$$p_{TIA} = 2\int_0^\pi p_{spin}(\theta) \cdot \cos(\theta) d\theta \qquad (6)$$

The probability of "spin" states to be in the TIS assembly can be calculated as (See Eq.(A1.8) in Appendix 1)

$$p_{TIS} = 1 - 2\int_0^\pi p_{spin}(\theta) \cdot \cos(\theta) d\theta \qquad (7)$$

The presented model of the TIA and TIS assemblies describes the magnetic properties of conduction electrons by means of a conversion of electrons between the assemblies. The electrons may be converted between the TIA and TIS assemblies by different physical mechanisms. In equilibrium all conduction electrons of a non-magnetic metal are in the TIS assembly. The total spin of TIS assembly is zero. This means that in equilibrium in a non-magnetic metal there is no spin accumulation. If there is a spin accumulation in the metal means that there are some electrons in the TIA assembly. The relaxation of the spin accumulation corresponds to a conversion of electrons from the TIA assembly into the TIS assembly. As will be shown in Chapter 5, the conduction electrons can be converted back from the TIS assembly into the TIA assembly under the applied magnetic field or due to exchange interactions with local d-electrons.

4. **Spin statistics.**

As was mentioned above, the energy distribution of electrons in both assemblies is described by the Fermi-Dirac statistic with the same Fermi energy. In this chapter the energy distributions of "spin" states, "full" states and "empty" state are calculated.

In the scattering event shown in Fig.2 as "scattering 2" an electron from a "full" state is scattered into an "empty" state. As a result, the number of "full" and "empty" states decreases and the number of "spin" states increases. In the scattering event shown in Fig.2 as "scattering 1" an electron from a "spin" state is scattered into another "spin" state. As a result of this scattering, the number of "full" and "empty" states increases and the number of "spin" states decreases. The average number of "spin", "full" and "empty" states in a metal is determined by the condition, that scatterings of "spin" states into "full" + "empty" states are balanced by the scatterings of "full" + "empty" states back into "spin" states.

At first, the spin distribution is calculated in the simplest case when all electrons are in the TIS assembly and there are no electrons in the TIA assembly. This corresponds to the case when there is no spin accumulation in the metal.

The energy distributions of electrons and holes in a metal are determined by the Fermi-Dirac statistic

$$F_{electrons}(E) = \frac{1}{1 + e^{\frac{E-E_F}{kT}}} \qquad (8)$$

$$F_{holes}(E) = \frac{1}{1+e^{-\frac{E-E_F}{kT}}} \qquad (9)$$

where $E_F$ is the Fermi energy.

The TIS assembly consists of "spin" states, "full" states and "empty" states. Each "spin" state is filled by one electron and one hole remains, each "full" state is filled by two electrons and "empty" state consists of two holes. These conditions are described as

$$F_{electrons} = F_{spin,TIS} + 2 \cdot F_{full}$$
$$F_{hole} = F_{spin,TIS} + 2 \cdot F_{empty} \qquad (10)$$

where $F_{spin,TIS}, F_{full}, F_{empty}$ are the energy distributions of "spin", "full" and "empty" states, respectively.

The result of an electron scattering out of a "full" state into an "empty" is two "spin" states. Therefore, the scattering event shown in Fig.2 as "scattering 2" causes an increase of the number of "spin" states with the rate

$$\left( \frac{\partial F_{spin,TIS}}{\partial t} \right)_{full \to spin} = 2 \cdot F_{full} \cdot F_{empty} \cdot p_{scattering} \qquad (11)$$

where $p_{scattering}$ is the probability of an electron scattering event per unit time.

A scattering of an electron out of a "spin" state into an empty place of another "spin" state (See Fig.3) reduces the number of "spin" states. According to Eq. (6) the probability of such a scattering event depends on the angle between the spin directions of "spin" states. If $F_{spin}(\varphi)$ is the distribution of "spin" states, which have the angle φ with the respect to the z-axis, from Eq.(6) the probability of scattering of "spin" states with angles φ₁ and φ₂ can be calculated as

$$p_{spin \to full}(\varphi_1, \varphi_2) = F_{spin}(\varphi_1) \cdot F_{spin}(\varphi_2) \cdot \sin^2\left((\varphi_1 - \varphi_2)/2\right) \qquad (12)$$

Since the spin of the electrons of the TIS assembly can have any direction with equal probability, the angular distribution $F_{spin}(\varphi)$ of "spin" states in TIS assembly can be calculated as

$$F_{spin}(\varphi) = 0.5 \cdot F_{spin} \cdot \sin(\varphi) \qquad (13)$$

where $F_{spin}$ is the number of "spin" sates at energy E.

From Eqs. (12) and (13), the reduction rate of "spin" states due to the scattering event shown in Fig.3 can be calculated by integrating Eq.(12) over all possible angles φ₁ and φ₂ as

$$\left(\frac{\partial F_{spin,TIS}}{\partial t}\right)_{spin \to full} = -2 \cdot p_{scattering} \cdot \int_0^{\pi} d\varphi_1 0.5 \cdot F_{spin} \cdot \sin(\varphi_1) \int_0^{\varphi_1} d\varphi_2 0.5 \cdot F_{spin} \cdot \sin(\varphi_2) \cdot \sin^2\left((\varphi_1 - \varphi_2)/2\right)$$
(16)

Integration of Eq.(16) gives

$$\left(\frac{\partial F_{spin,TIS}}{\partial t}\right)_{spin \to full} = -0.5 \cdot p_{scattering} \cdot F_{spin}^2 \left(1 - \frac{\pi^2}{16}\right) \approx 0.1916 \cdot p_{scattering} \cdot F_{spin}^2 \quad (17)$$

In equilibrium, the conversion rate of "spin" states into "full" states (Eq.17) is equal to the rate of the back conversion Eq.(13)

$$\left(\frac{\partial F_{spin,TIS}}{\partial t}\right)_{full \to spin} + \left(\frac{\partial F_{spin,TIS}}{\partial t}\right)_{spin \to full} = 0 \quad (18)$$

Substituting Eq. (11) and Eq. (17) into Eq. (18) gives

$$0.1916 \cdot p_{scattering} \cdot F_{spin,TIS}^2 = 2 \cdot F_{full} \cdot F_{empty} \cdot p_{scattering} \quad (19)$$

Solving Eqs.(19) and (10) gives

$$F_{spin,TIS} = \frac{1}{0.6168} \left( \frac{F_{electrons} + F_{spin}}{2} - \sqrt{\left(\frac{F_{electrons} + F_{spin}}{2}\right)^2 - 0.6168 \cdot F_{electrons} F_{hole}} \right) \quad (20)$$

Substituting (8) and (9) into (20) gives the distribution of "spin" states in the TIS assembly as

$$F_{spin,TIS}(E) = \frac{1}{1.2336} \left( 1 - \sqrt{1 - \frac{1.2336}{1 + \cosh\left(\frac{E - E_F}{kT}\right)}} \right) \quad (21)$$

Figure 5(a) shows the distribution of "spin", "full" and "empty" states calculated from Eqs. (20). The "spin" states are mainly distributed near the Fermi energy. The "full" states are mainly distributed at lower energies and "empty" states are mainly distributed at higher energies. It should be noticed that for energies larger than ~2 kT above the Fermi energy the number of "spin" states is significantly greater than the number of "full" states. This means that in this case almost all electrons are in "spin" states. Similar, for energies smaller than ~2 kT below the Fermi energy, all holes are in "spin" states.

The total number of "spin" states in the TIS assembly can be calculated as

$$n_{spin} = \int_{-\infty}^{\infty} dE \cdot D(E) \cdot F_{spin}(E) \quad (22)$$

where D(E) is the density of states.

In the case of a metal, which has a near constant density of states near the Fermi energy, the total number of "spin" states is calculated using Eqs. (21) and (22) as

$$n_{spin} \approx D(E) \cdot kT \cdot 1.1428 \quad (23)$$

In the following the energy distribution for the case when there are electrons in both TIA and TIS assemblies is calculated.

The spin life time of conduction electrons in a solid can not be infinite. There is always a conversion of electrons of the TIA assembly into the TIS assembly due to the different spin relaxation mechanisms. This conversion is characterized by the spin relaxation rate:

$$\left(\frac{\partial n_{TIA}(E)}{\partial t}\right)_{TIA \to TIS} = -\frac{n_{TIA}(E)}{\tau_{spin}} \quad (24)$$

where $n_{TIA}(E)$ is the number of electrons in TIA assembly at energy E and $\tau_{spin}$ is the spin life time.

In order for electrons to be in the TIA assembly, there should be some mechanism, which converts the electrons back from the TIS assembly into the TIA assembly. As will be shown below, such a conversion can be induced either by a magnetic field or an effective exchange field due to the exchange interaction or due to an absorption of circularly polarized light. This conversion can be expressed as

$$\left(\frac{\partial n_{TIA}(E)}{\partial t}\right)_{TIS \to TIA} = \frac{n_{spin,TIS}(E)}{\tau_{conversion}} \quad (25)$$

where $n_{TIS}(E)$ is the number of "spin" states in the TIS assembly at energy E and $\tau_{conversion}$ is the effective time of the conversion.

In equilibrium the conversion rates (24) and (25) should be the same. This condition gives a ratio of "spin" states in the TIA assembly to the total number of the spin states as

$$x(E) = \frac{n_{TIA}(E)}{n_{TIA}(E) + n_{spin,TIS}(E)} = \frac{\tau_{conversion}}{\tau_{conversion} + \tau_{spin}} \quad (26)$$

It can be assumed that x(E) only weakly depends on the electron energy E. In this case x is defined as the spin polarization of conduction electrons.

For the case when there are electrons in both TIA and TIS assemblies, Eqs. (10) should be modified as

$$F_{electrons} = F_{spin,TIA} + F_{spin,TIS} + 2 \cdot F_{full}$$
$$F_{hole} = F_{spin,TIA} + F_{spin,TIS} + 2 \cdot F_{empty} \quad (27)$$

where $F_{spin,TIA}, F_{spin,TIS}$ are the energy distributions of "spin" states in the TIA and TIS assemblies, respectively.

From Eq (26), the energy distributions of TIA and TIS assemblies are related as

$$F_{spin,TIA} = F_{spin,TIS} \frac{x}{1-x} \qquad (28)$$

Solving Eqs. (27), (28) and (19) gives the distribution of "spin" state in the TIS assembly and the TIA assembly as

$$F_{spin,TIS} = \frac{1-x}{2 \cdot \left(1 - 0.3832(1-x)^2\right)} \left(1 - \sqrt{1 - \frac{2 \cdot \left(1 - 0.3832(1-x)^2\right)}{1 + \cosh\left(\frac{E-E_F}{kT}\right)}}\right) \qquad (29)$$

$$F_{spin,TIA} = \frac{x}{2 \cdot \left(1 - 0.3832(1-x)^2\right)} \left(1 - \sqrt{1 - \frac{2 \cdot \left(1 - 0.3832(1-x)^2\right)}{1 + \cosh\left(\frac{E-E_F}{kT}\right)}}\right) \qquad (30)$$

Figure 5(b) shows the energy distribution of "spin" states of the TIA assembly and "spin", "full" and "empty states of the TIS assembly in the case when the spin polarization of conduction electrons is 75%.

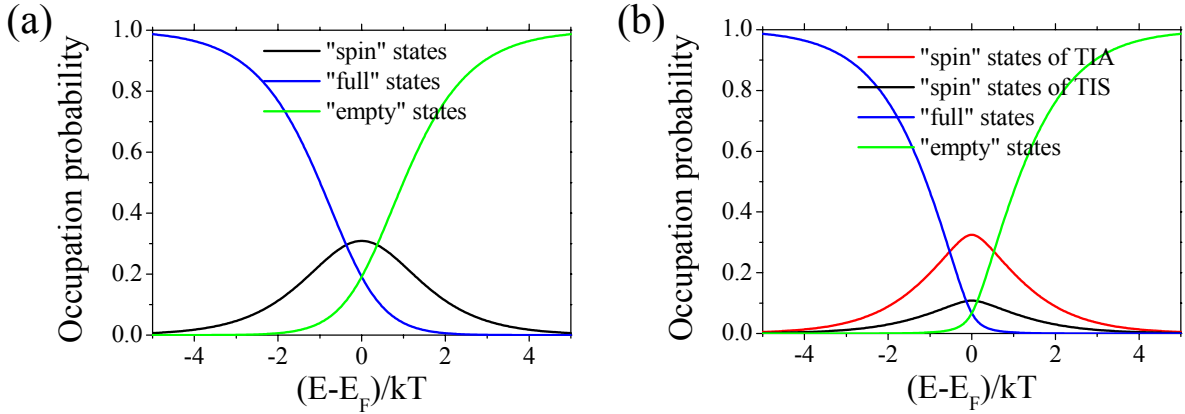

*Fig.5 The energy distribution of "spin", "full" and "empty" states in cases (a) no spin accumulation (b) spin polarization = 75%*

For the electron energies significantly above the Fermi energy

$$\frac{E - E_F}{kT} \gg 1 \qquad (31)$$

Eqs. (29) and (30) are simplified to

$$F_{spin,TIS} \approx (1-x) e^{-\frac{E-E_F}{kT}}$$

$$F_{spin,TIA} \approx x \cdot e^{-\frac{E-E_F}{kT}} \qquad (32)$$

As was mentioned above in case of energies, which satisfy condition (31), most of the electrons are in "spin" states. Therefore, the energy distribution for the "spin" states in this case should be the same as energy distribution of the electrons. Since under the conditions (31) the energy distribution of electrons is the Boltzmann distribution, the distribution of "spin" states in this case is described by the Boltzmann statistics (Eq. (32)) as well.

The importance of the spin statistic for the description of spin transport should be mentioned. The spin statistic should be used in the Boltzmann transport equation, which describes the spin transport. This study will be published elsewhere.

**5. Spin accumulation induced by magnetic field. Conversion of TIS into TIA induced by magnetic field**

The behavior of the TIA and TIS assemblies is different in a magnetic field. The TIA assembly has a spin direction and there is a precession of the total spin of the TIA assembly around direction of the magnetic field. The TIS assembly has zero total spin and it does not precess around the magnetic field. However, the TIS assembly contains " spin" states, whose spins precess around the direction of the magnetic field. This precession itself does not affect the assembly. Even in a magnetic field the probability to find a "spin" state in the TIS assembly is equal for all spin directions. However, the damping of the precession has a significant influence on the TIS assembly. It causes a conversion of electrons from the TIS assembly into the TIA assembly.

During the spin precession, the direction of spin slowly turns in the direction of the magnetic field. This effect is called precession damping. The spin precession and precession damping is described by the Landau-Lifshiz equation:

$$\frac{d\vec{M}_{el}}{dt} = -\gamma \vec{M}_{el} \times \vec{H} + \lambda \vec{M}_{el} \times \left(\vec{M}_{el} \times \vec{H}\right) \qquad (34)$$

where $\gamma$ is the electron gyromagnetic ratio, $\lambda$ is a phenomenological damping parameter, $\vec{M}_{el} = -g \cdot \mu_B \cdot \vec{S}$ is the magnetic moment of a conductive electron, g is the g-factor and $\mu_B$ is the Bohr magneton.

In the case when the magnetic field is applied along the z-axis, a solution of the Landau-Lifshiz equation (34) can be found (See Appendix 2). In this solution the z-component of the spin vector is described as

$$S_z = \frac{1}{2}\cos(\theta(t)) \qquad (35)$$

where $\theta$ is the angle between the direction of the magnetic field and the spin direction. $\theta$ can be found from the differential equation:

$$V_\theta(t) = \frac{d\theta(t)}{dt} = -\frac{1}{t_\lambda}\sin(\theta(t)) \qquad (36)$$

where

$$t_\lambda = \frac{2}{\lambda \cdot g \cdot \mu_B \cdot H_z} \qquad (37)$$

is defined as the precession damping time and $V_\theta(\theta)$ is the angular speed of spin rotation towards the magnetic field.

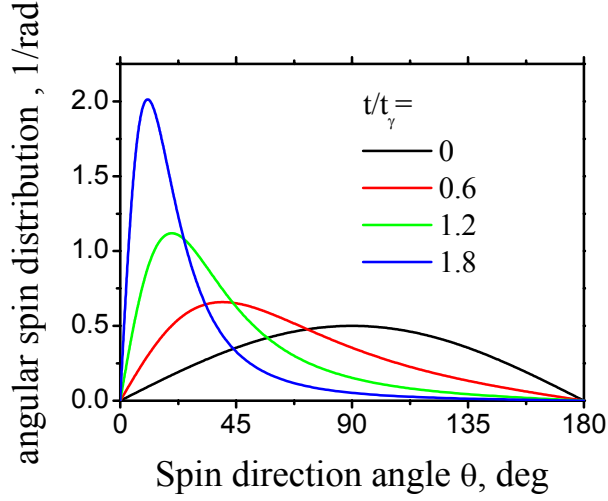

*Fig.6 Angular spin distribution of electrons in the TIS assembly for different time intervals after a magnetic field has been turned on. The distribution at time t=0 corresponds to the homogeneous distribution of the TIS assembly in the absence of a magnetic field.*

In the TIS assembly a "spin" state has any spin direction with equal probability. Under a magnetic field, spins precess around the direction of the magnetic field. Also, spins turn toward the magnetic field, because of the precession damping. This changes the angular distribution of spin directions, increasing the probability of finding "spin" states with a smaller angle with respect to the magnetic field direction.

In the following the temporal evolution of the spin angular distribution for the TIS assembly is calculated. Assuming that we know the angular distribution of spin $\rho_{spin}(\theta,t)$ at time t, we can calculate the angular distribution at time t+dt. For example, if some states had a spin angle in the interval $(\theta_0, \theta_0 + d\theta_0)$ at time t, due to precession damping their spins turn toward the z-axis and at time t+dt their spin angle will be in the interval $(\theta_1, \theta_1 + d\theta_1)$. Assuming that time dt is sufficiently short, $\theta_1$ and $\theta_1 + d\theta_1$ can be calculated as

$$\theta_1 = \theta_0 + V_\theta(\theta_0) \cdot dt$$

$$\theta_1 + d\theta_1 = \theta_0 + d\theta_0 + V_\theta(\theta_0 + d\theta_0) \cdot dt = \theta_0 + d\theta_0 + V_\theta(t,\theta_0) \cdot dt + \frac{\partial V_\theta(t,\theta_0)}{\partial \theta_0} \cdot d\theta_0 \cdot dt \quad (38)$$

The precession damping does not change the number of spins, it only changes their directions. Therefore, the number of "spin" states is the same in the interval $(\theta_1, \theta_1 + d\theta_1)$ as in the interval $(\theta_0, \theta_0 + d\theta_0)$. Using this fact and Eq (38), the angular distribution of spin directions at time t+dt can be calculated as

$$\rho_{spin}(\theta_1, t+dt) = \frac{p_{spin}(\theta_1, t+dt)}{[\theta_1 + d\theta_1] - \theta_1} = \frac{p_{spin}(\theta_0, t)}{d\theta_0 + \frac{\partial V_\theta(t, \theta_0)}{\partial \theta_0} \cdot d\theta_0 \cdot t} = \frac{\rho_{spin}(\theta_0, t)}{1 + \frac{\partial V_\theta(t, \theta_0)}{\partial \theta_0} \cdot dt} \quad (39)$$

where

$\rho_{spin}(\theta_0, t) = \frac{p_{spin}(\theta_0, t)}{[\theta_0 + d\theta_0] - \theta_0}$ is the angular spin distribution at time t

and $p_{spin}(\theta_0, t) = p_{spin}(\theta_1, t+dt)$ is the probability for a "spin" state to have spin direction in the interval $(\theta_1, \theta_1 + d\theta_1)$ at time t+dt, which is equal to the probability of having spin direction in the interval $(\theta_0, \theta_0 + d\theta_0)$ at time t

Substituting Eq.(38) into Eq.(39) gives

$$\rho_{spin}(\theta_0 + V_\theta(\theta_0) \cdot dt, t+dt) = \frac{\rho_{spin}(\theta_0, t)}{1 + \frac{\partial V_\theta(t, \theta_0)}{\partial \theta_0} \cdot dt} \quad (40)$$

Substituting Eq (37) into Eq. (40) gives an equation, which describes the temporal evolution of the angular distribution of spin directions as:

$$\rho_{spin}\left(\theta_0 - \frac{t}{t_\lambda}\sin(\theta_0) \cdot dt, t+dt\right) = \frac{\rho_{spin}(\theta_0, t)}{\left(1 - \frac{dt}{t_\lambda}\cos(\theta_0)\right)} \quad (41)$$

In order to solve Eq. (41), an initial condition should be used. For example, let us assume that at time t=0 a magnetic field has been applied, and the spins starts to rotate toward the magnetic field. At the time t=0 before the magnetic field has been applied, the "spin" states of the TIS assembly have equal probability for their spins to be in any direction. Therefore, the angular spin distribution at time t=0 can be described as

$$\rho_{spin}(\theta, t=0) = 0.5 \cdot \sin(\theta) \quad (42)$$

It should be noticed that the integration over all angles $\int_{\theta=0}^{\pi} \rho_{spin}(\theta, t) d\theta$ gives the probability for a "spin" state to be in TIS assembly and it should be equal to one at any time, because the precession damping does not change the total number of spins.

Fig.6 shows the time evolution of the spin angular density obtained by solving numerically Eq (41) using the initial condition (42). For time longer than $t_\lambda$, spins of the TIS assembly are mainly directed along the magnetic field. This could only happen in the case when there are no scatterings between electron states. The scattering event shown in Fig. 3 always converts electrons into one TIS assembly, where spins are in all directions with equal probability, and one TIA assembly, where all spins are in the same direction. Therefore, the scattering event shown in Fig. 3 does not give spins of the TIS assembly sufficient time to turn fully toward the magnetic field. Even after a slight rotation toward the magnetic field, the electrons are scattered back into the assembly, where spin is distributed equally in all directions. The access of "spin" states, which are directed towards the magnetic field, is converted into the TIA assembly.

The scattering event shown in Fig. 3 is rather frequent. The precession damping time $t_\lambda$ is inversely proportional to the intensity of the magnetic field (Eq. 37). In the case of a small or moderate magnetic field, $t_\lambda$ should be significantly longer than the scattering time $t_{scattering}$

$$t_{scattering} \ll t_\gamma \tag{43}$$

This means that the rotation angle of spins towards the magnetic field is very small between scatterings. In this case it is possible to solve analytically Eq. (41) and to find the conversion rate.

Under condition (43) the angular spin distribution can be expressed as

$$\rho_{spin}(\theta,t) = \rho_{spin,TIS}(\theta) + \Delta\rho_{spin}(\theta,t) \cdot \frac{t}{t_\lambda} \tag{44}$$

where $\rho_{spin,TIS}(\theta)$ is the angular spin distribution of the TIS assembly (Eq. (42)). Using Eq. (44), the solution of Eq. (41) is found as

$$\Delta\rho_{spin}(\theta) = n_{spin,TIS} \cdot \sin(\theta)\cos(\theta) \tag{45}$$

The scatterings quickly convert any inhomogeneous distribution of "spin" states into one TIA and one TIS assembly. The number of "spin" states converted into the TIA assembly can be calculated by Eq.(6). Substituting Eqs. (44) into Eq.(33) and noticing that the contribution into the integral from $\rho_{spin,TIS}(\theta)$ is zero, the probability for "spin" states to be in the TIA assembly can be calculated as

$$p_{TIA}(t) = 2\int_0^\pi \Delta\rho_{spin}(\theta,t) \cdot \cos(\theta) d\theta \tag{46}$$

Integrating (46) gives the rate of conversion of "spin" states from the TIA assembly into the TIA assembly as

$$\frac{\partial n_{TIA}}{\partial t} = \frac{4}{3} \frac{n_{spin,TIS}}{t_\lambda} \tag{47}$$

where $n_{TIA}, n_{spin,TIS}$ are the numbers of "spin" states in the TIA and TIS assemblies, respectively.

The spin damping time $t_\lambda$ decreases when the intensity of the magnetic field increases. In the case of a large magnetic field, the condition (43) may not be satisfied, in this case the conversion rate can be calculated as

$$\frac{\partial n_{TIA}}{\partial t} = k\left(\frac{t_{scattering}}{t_\lambda}\right) \cdot \frac{n_{spin,TIS}}{t_\lambda} \qquad (48)$$

where $k\left(\frac{t_{scattering}}{t_\lambda}\right)$ is the conversion-efficiency coefficient, which depends on the ratio of the scattering time to the spin damping time.

Figure 7 shows the conversion-efficiency coefficient k, which was evaluated from a numerical solution of Eq.(41). The conversion-efficiency coefficient k becomes smaller than 4/3 for a stronger magnetic field.

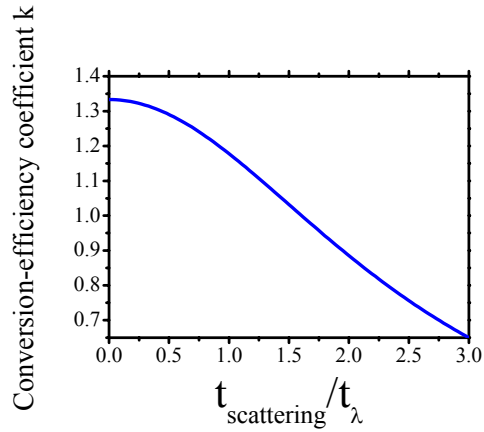

*Fig.7 Conversion-efficiency coefficient k as the function of the ratio of the scattering time to the spin damping time. In the case of a short scattering time or a weak magnetic field, k=4/3.*

In a metal, in which the d-orbitals are partially filled and which has local d-electrons with non-zero spin and the conductive sp-electrons, there could be a significant exchange interaction between the d-electrons and the conductive electrons. In the case when spins of all d-electrons directed along the same direction (for example, in the case of a ferromagnetic metal), the exchange interaction causes a conversion of "spin" states from the TIS assembly into the TIA assembly.

The exchange interaction causes a precession of the spins of the conduction electrons around the direction of the local d-electrons. Because of the damping of this precession, "spin" states of the TIS assembly turn their spin toward the spin direction of the d-electrons. This causes a conversion of "spin" states from the TIS assembly into TIA assembly. Similar to Eq.(48) the conversion rate due to the exchange interaction can be calculated as

$$\frac{\partial n_{TIA}}{\partial t} = k\left(\frac{t_{scattering}}{t_{\lambda,exchange}}\right) \cdot \frac{n_{spin,TIS}}{t_{\lambda,exchange}} \qquad (49)$$

where

$$t_{\lambda,exchange} = \frac{2}{\lambda_{exchange} \cdot g \cdot \mu_B \cdot H_{exch,eff}} \qquad (50)$$

$H_{exch,eff}$ is the effective magnetic field of the exchange interaction and $\lambda_{exchange}$ is a phenomenological damping parameter for the exchange interaction.

It should be noted that in the case when spins of the local d-electrons and the conduction sp-electrons are comparable, there are precessions of both the local d-electrons and the conduction electrons around a common axis.

## 6. Pauli paramagnetism in non-magnetic metals

Pauli paramagnetism is the paramagnetism in metals induced by conductive electrons. The Pauli theory explains this paramagnetism based on the classical model of the spin-up/spin-down bands as follows. In the absence of any magnetic field, there is equal number of electrons in the spin-up and spin-down bands and the total magnetic moment of conduction electrons is zero. In a magnetic field the spin-up and spin-down electrons have different energy. In order to make the Fermi energy is equal for both bands, after a magnetic field is applied, some electrons from the spin-down band are flipped into the spin-up band. Therefore, the spin-up band will be filled by a greater number of electrons than the spin-down band. Due to the difference in the number of spin-up and spin-down electrons, the net magnetization of conduction electrons becomes non-zero[16].

The Pauli paramagnetism is calculated from model of TIS/TIA assemblies differently. In the absence of a magnetic field all conduction electrons of a non-magnetic metal are in the TIS assembly. Even though there are "spin" states in the TIA assembly, the total magnetic moment of the TIS assembly is zero, because this assembly is time-inverse symmetrical. The magnetic moment is time-inverse asymmetrical and only time-inverse asymmetrical objects may have non-zero magnetic moment. Therefore, the total magnetic moment of an electron gas may be non-zero only in the case when some electrons are in the TIA assembly.

As was explained in the previous chapter, under a magnetic field there is a conversion of the "spin" states from the TIS assembly into the TIA assembly at the rate (See Eq. (48))

$$\left(\frac{\partial n_{TIA}}{\partial t}\right)_{TIS \to TIA} = k \cdot \frac{n_{TIS}}{t_\lambda} \quad (51)$$

Also, there is a back conversion of the "spin" states from the TIA assembly into the TIS assembly due to spin relaxation mechanisms. The rate of this conversion is proportional to the number of electrons in the TIA assembly

$$\left(\frac{\partial n_{TIA}}{\partial t}\right)_{TIA \to TIS} = -\frac{n_{TIA}}{\tau_{spin}} \quad (52)$$

where $\tau_{spin}$ is the spin life time.

In equilibrium the rate of conversion of electrons from the TIS assembly into the TIA assembly should be equal to the rate of reverse conversion

$$\left(\frac{\partial n_{TIA}}{\partial t}\right)_{TIS \to TIA} + \left(\frac{\partial n_{TIA}}{\partial t}\right)_{TIA \to TIS} = 0 \quad (53)$$

From Eq. (53) the equilibrium number of electrons in the TIA assembly can be calculated as

$$n_{TIA} = k \frac{\tau_{spin}}{t_\lambda} n_{TIS} \qquad (54)$$

Therefore, a magnetic field induces spin accumulation in a non-magnetic metal and the spin polarization of the non-magnetic metal becomes non-zero. Only the electrons of the TIA assembly contributes to the induced magnetization. The magnetic moment per "spin" state of the TIA assembly is

$$\mu_S = \frac{g \cdot \mu_B}{2} \qquad (55)$$

where g is the g-factor, $\mu_B$ is the Bohr magneton.

The induced magnetic moment of an electron gas is equal to the total magnetic moment of TIA assembly and it can be calculated as

$$M_{induced} = \frac{g \cdot \mu_B}{2} n_{TIA} = \frac{(g \cdot \mu_B)^2}{4} k \cdot \tau_{spin} \lambda \cdot n_{TIS} \cdot H \qquad (56)$$

The number of "spin" states in the TIS assembly $n_{TIS}$ should be calculated using the spin statistics Eqs.(29),(30)

In case of a weak magnetic field (See condition (43)), neither coefficient k nor $n_{TIS}$ depends on the magnetic field and the induced magnetization is linearly proportional to the magnetic field. This means that in this case the permeability does not depend on the intensity of the magnetic field.

For a stronger magnetic field the permeability decreases when the magnetic field increases. It is because the coefficient k decreases (See Fig.6) and the number of "spin" states in the TIS assembly decreases. The TIS assembly has a smaller number of "spin" states, because some "spin" states have been converted into the TIA assembly.

In case of an even larger magnetic field, nearly all "spin" states may be converted into the TIA assembly (the metal is near 100 % spin polarized) and the induced magnetization (Eq. (56)) saturates at a value

$$M = \frac{g \cdot \mu_B}{2} n_{TIA} \approx \frac{g \cdot \mu_B}{2} n_{spin} \qquad (57)$$

where $n_{spin}$ is the total number of spin states.

**7. Ferromagnetic metals**

The magnetic properties of an electron gas in a ferromagnetic metal have been described by the classical Stoner model[20]. In the Stoner model it is assumed that the conduction electrons in a ferromagnetic metal are divided into two groups: the majority band and the minority band. In each band, spins of all electrons are aligned either parallel or anti parallel to the magnetization direction of the ferromagnetic metal. The density of states for the majority and minority bands was assumed to be significantly different. It was assumed that there is only a weak interaction between the electrons of different bands so that it is possible to assign the chemical potential individually for electrons of each band. Because of the spin-dependent density of states, the conductivity of a ferromagnetic metal becomes spin-dependent, which leads to several interesting effects for spin and charge transport such as a charge accumulation, a shortening of spin diffusion length and a blocking of spin diffusion.

The scattering event shown in Fig.3 is frequent. If electrons would only be in the majority and minority bands, this scattering event would mix electrons from both bands within a very short time. That fact strongly limits the validity of the classical Stoner model. The model of TIS/TIA takes into account the spin rotation due to the scattering event shown in Fig.3 as follows.

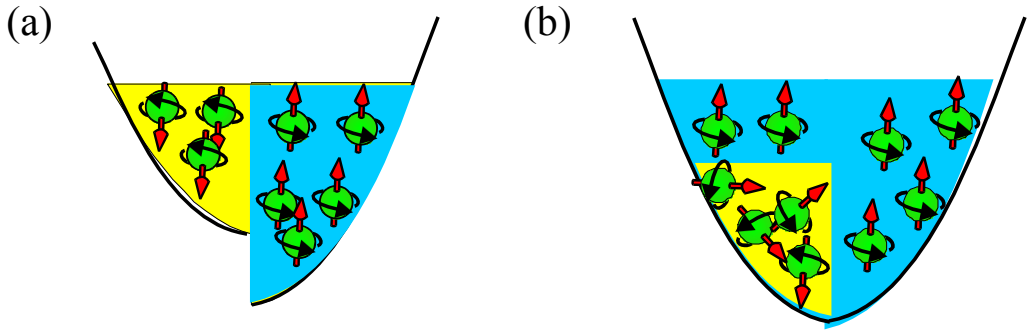

*Fig.8. The distribution of spin directions in a ferromagnetic metal. (a) According to the classical model of spin-up/spin-down bands. The blue area corresponds to the majority band. The yellow area corresponds to the minority band. (b) According to the model of TIS/TIA assemblies. The blue area corresponds to the TIA assembly. The yellow area corresponds to the TIS assembly.*

In equilibrium the conduction electrons in a ferromagnetic metal are in one TIA assembly and in one TIS assembly. The spin direction of the TIA assembly is aligned along the metal magnetization. The "spin" states of the TIS assembly have equal probability to be directed in any direction. Figure 8 compares the distribution of spins in a ferromagnetic metal according to the model of spin-up/spin-down bands and to the distribution according to the model of TIS/TIA assemblies.

In a ferromagnetic metal the d-orbitals are partially filled and electrons can be divided into local d-electrons with non-zero spin and conductive sp-electrons. There is an exchange interaction between d-electrons, which aligns spins of all d-electrons in one direction. Also, there is an exchange interaction between the local d-electrons and the conduction sp-electrons, which leads to the conversion of "spin" states from the TIS assembly into the TIA assembly. In addition, there is a back conversion from the TIA assembly into the TIS assembly, because of a finite spin life time. In equilibrium the conversion of electrons from the TIS to the TIA assembly is balanced by the back conversion.

The conversion rate from the TIS assembly to the TIA assembly, which originates from the exchange interaction between sp- and d-electrons, is proportional to the number of "spin" states in the TIS assembly (Eq.(49))

$$\left(\frac{\partial n_{TIA}}{\partial t}\right)_{TIS \to TIA} = k \cdot \frac{n_{TIS}}{t_{\lambda, exchange}} \qquad (58)$$

The conversion rate from the TIA assembly to the TIS assembly, which has contributions from different spin relaxations mechanisms, is proportional to the number of electrons in the TIA assembly

$$\left(\frac{\partial n_{TIA}}{\partial t}\right)_{TIA \to TIS} = -\frac{n_{TIA}}{\tau_{spin}}. \tag{59}$$

In equilibrium both rates should be the same

$$\left(\frac{\partial n_{TIA}}{\partial t}\right)_{TIA \to TIS} + \left(\frac{\partial n_{TIA}}{\partial t}\right)_{TIS \to TIA} = 0 \tag{60}$$

From Eq. (60) the spin polarization of conductive electrons in a ferromagnetic metal can be calculated as

$$x = \frac{n_{TIA}}{n_{TIS} + n_{TIA}} = \frac{k \cdot \tau_{spin}}{t_{\lambda,exchange} + k \cdot \tau_{spin}} \tag{61}$$

It should be noted that the model of TIS/TIA assemblies does not necessarily require a difference of the densities of states for electrons with spin parallel and anti-parallel to the metal magnetization.

**8. Spin torque. The interaction of two TIA assemblies**

The spin torque is the torque, which is rotates the spin direction of the TIA assembly toward the spin direction of electrons of another TIA assembly injected into or generated in a metal.

When there is a spin accumulation in a metal and a small amount of electrons of another spin direction is injected into the metal or converted from the TIS assembly (See Chapter 5), the spin direction of the spin accumulation is rotating toward spin direction of the injected (converted) electrons. Therefore, the injected (converted) electrons cause a torque, which acts on the spin accumulated electrons. It should be noted that the spin torque is always accompanied by an additional spin relaxation. This means that during the interaction of the TIA assemblies some electrons from these assemblies are converted into the TIS assembly.

The spin torque occurs because of the interaction of two TIA assemblies. Some features of this interaction are studied below.

The interaction of two TIA assemblies significantly depends on the relative number of electrons in each assembly. As was shown in Chapter 4, the scattering event shown in Fig. 3 causes electrons of different TIA assemblies to combine into one TIA assembly. In the case of the interaction of two TIA assemblies with the same number of electrons, one scattering event shown in Fig.3 between electrons of the TIA assemblies is sufficient to combine the assemblies. In the case of the interaction of two TIA assemblies with a different number of electrons, it takes several scatterings until the two TIA assemblies relax into one TIA assembly. For example, let us consider the interaction of TIA1 and TIA2 assemblies when the number of electrons in TIA1 is greater than the number of electrons in TIA2. The spin direction of TIA1 is along the z-axis and the spin direction of TIA2 has angle ϕ with respect to the z-axis. After the first scattering event shown in Fig.3 the spin direction of TIA2 will have the angle ϕ/2 with respect to the z-axis, the spin direction of TIA1 will not change. The number of electrons in TIA1, TIA2 and TIS assemblies after the first scatterings will be

$$n_{TIA1,1} = n_{TIA1} - n_{TIA1}$$
$$n_{TIA2,1} = 2 \cdot n_{TIA2} \cos^2(\phi/2)$$
$$n_{TIS} = 2 \cdot n_{TIA2} \sin^2(\phi/2)$$
(62)

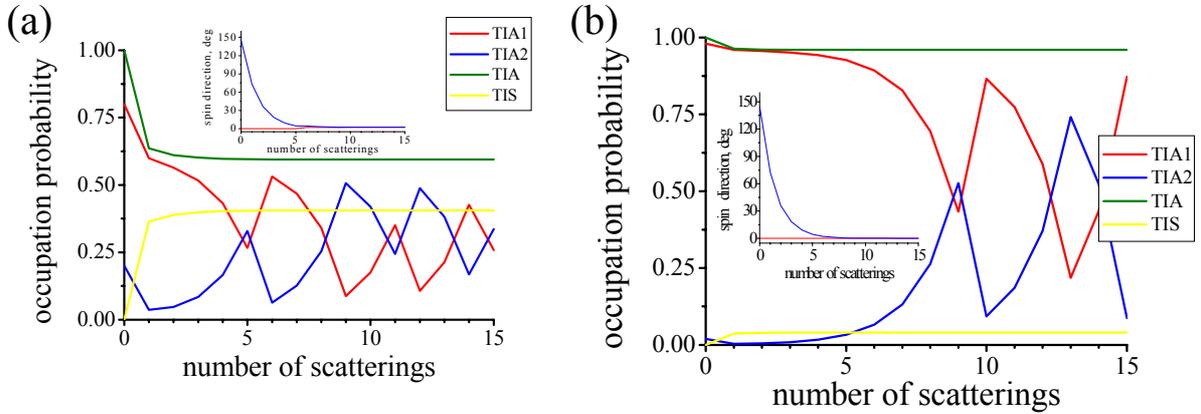

*Fig.9 Interaction of two TIA assemblies. Occupation probabilities of TIA1 assembly (red line), TIA2 assembly (blue line), either of the TIA assemblies (green line) and the TIS assembly (yellow line). Initial conditions: (a) TIA1 assembly: probability=0.8 angle= 0 deg and TIA2 assembly: probability=0.2 angle= 145 deg; (b) TIA1 assembly: probability=0.98 angle= 0 deg and TIA2 assembly: probability=0.02 angle= 145 deg*

After the second scattering event shown in Fig.3, the angle between the spin directions of TIA assemblies will be ϕ/4. In the case when the number of electrons in TIA2 assembly is greater than in TIA1 assembly, the number of electrons in TIA1, TIA2 and TIS assemblies will be

$$n_{TIA1,2} = 2 \cdot n_{TIA1,1} \cos^2(\phi/4)$$
$$n_{TIA2,2} = n_{TIA2,1} - n_{TIA1,1}$$
$$n_{TIS} = 2 \cdot n_{TIA2} \sin^2(\phi/2) + 2 \cdot n_{TIA2,0} \sin^2(\phi/4)$$
(63)

Otherwise, when number of electrons in TIA1 is greater than in TIA2, the spin direction of TIA2 assembly rotates.

Therefore, after each scattering event shown in Fig.3, the angle between the two TIA assemblies is reduced by a factor of 2. The spin direction of the assembly, in which there were fewer electrons, rotates and the number of electrons in this assembly increases. The spin direction of the assembly, in which there were more electrons, does not rotate and the number of electrons in this assembly decreases.

It should be noticed that the probability of the scattering event shown in Fig.3 is not constant, but it is proportional to the number of electrons in the TIA1 and TIA2 assemblies. The probability of one scattering event shown in Fig.3 can be calculated as

$$p_{scat5} = p_{scat5,0} \cdot \frac{n_{TIA1} \cdot n_{TIA2}}{n_{TIA1} + n_{TIA2} + n_{TIS}} \tag{64}$$

where $n_{TIA1}, n_{TIA2}, n_{TIS}$ are the number of "spin" states in the TIA1, TIA2 and TIS assemblies, respectively. $p_{scat5,0}$ is the probability of a single scattering event shown in Fig.3.

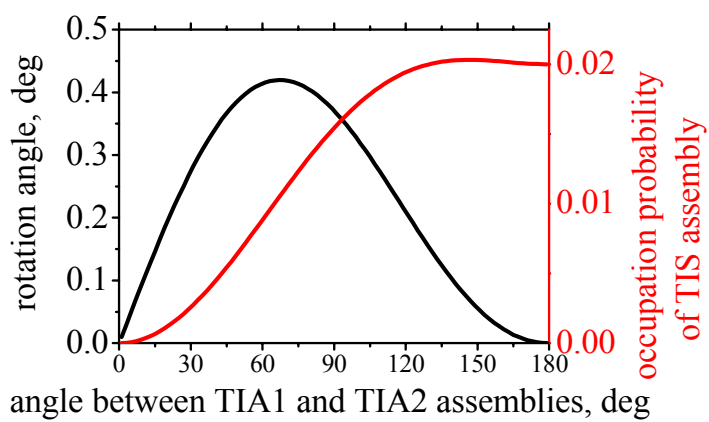

*Fig.10 Interaction of two TIA assemblies. The rotation angle and the probability for an electron to be converted into the TIS assembly as a function of the initial angle between the TIA1 and TIA2 assemblies. Initially 99 % of the electrons were in TIA1 assembly and 1 % of the electrons were in TIA2 assembly. Initially there were no electrons in TIS assembly.*

The interaction of two TIA assemblies was simulated numerically by the method described above. Figure 9(a) shows the case of the interaction of the TIA1 and TIA2 assemblies when initially 80 % of the conduction electrons are in the TIA1 assembly, 20 % of the conductive electrons are in the TIA2 assembly and there are no electrons in the TIS assembly. Figure 9(b) shows a similar case of the interaction of the TIA1 and TIA2 assemblies. However, initially 98 % of the conduction electrons are in the TIA1 assembly and 2 % of the conductive electrons are in the TIA2 assembly. The initial angle between spin directions of the TIA assemblies was 145 deg in both cases. After each scattering the angle between the TIA assembles decreases and some electrons are converted into the TIS assembly. During the first 3-4 scatterings there is a significant conversion of the electrons from the TIA1 assembly into TIS assembly (spin relaxation), the number of electrons in the TIA1 assembly increases and the number of electrons in the TIA2 assembly decreases. Only the spin direction of the TIA1 assembly changes during the first scatterings. However, after 5-7 scatterings the number of electrons in the TIA assemblies becomes comparable and the spin direction of both assemblies rotates. After 10-15 scatterings the angle between the TIA assemblies becomes very small and the two assemblies can be considered as one TIA assembly.

Figure 10 shows the rotation angle and the amount of electrons converted into the TIS assembly as a result of the interactions of the TIA1 and TIA2 assemblies, for the case when 99 % of conduction electrons are in TIA1 and 1 % of conduction electrons are in TIA2. There is no rotation in the cases when the spin directions of TIA1 and TIA2 are parallel or antiparallel. The rotation is largest in the case when the angle between the spin directions of the TIA assemblies is around 67 degrees.

The number of electrons converted into the TIS assembly increases when the angle between the assemblies increases. When the angles is 120 degrees or larger, the number of converted electrons is around 2%. That is twice the initial amount of TIA2 electrons. For angles smaller than 67 degrees, the amount of converted electrons is smaller than the initial amount of TIA2 electrons. Therefore, the number of electrons in the resulting TIA assembly is larger than the initial number in TIA1. This means that the number of electrons in the TIA assembly increases because of the injection. In contrast, in the case of angles greater than 67 degrees, the number of electrons in the TIA assembly decreases due to the injection.

In the case when a small amount of electrons of TIA2 assembly is continuously injected (or converted from TIS assembly) at a rate $\frac{\partial n_{TIA2}}{\partial t}$ into the region where there are some electrons in the TIA1 assembly, the electrons of TIA1 experience a spin torque, which can be calculated as

$$\frac{\partial \vec{s}_{TIA1}}{\partial t} = \frac{A(\phi)}{n_{TIA1}} \frac{\partial n_{TIA2}}{\partial t} \cdot \vec{s}_{TIA1} \times [\vec{s}_{TIA1} \times \vec{s}_{TIA2}] \qquad (65)$$

where $\vec{s}_{TIA1}$ and $\vec{s}_{TIA2}$ are unit vectors directed along the spin direction of the TIA1 and TIA2 assemblies and $n_{TIA1}$ is the number of electrons in the TIA1 assembly.
Also, the injection (the conversion) causes a conversion of electrons from the TIA assembly to the TIS assembly at a rate

$$\frac{\partial n_{TIS}}{\partial t} = B(\phi) \frac{\partial n_{TIA2}}{\partial t} (1 - \vec{s}_{TIA1} \cdot \vec{s}_{TIA2}) \qquad (66)$$

The coefficients $A(\phi)$ and $B(\phi)$ depend on the angle $\phi$ between the assemblies. These coefficients may be found by the above-discussed method of calculations of the interaction of two TIA assemblies, assuming that the injection rate is small. Figure 11 shows the spin torque coefficient A and the TIS conversion coefficient B as the functions of the angle between the spin directions of the TIA1 and TIA2 assemblies. For small angles the coefficient A equals ~57 deg and the coefficient B equals ~2.

**9. Spin-torque current.**

The spin-torque current is a current flowing between regions, which have different spin directions of the TIA assembly. The spin-torque current induces a spin torque on the electrons of the TIA assembly. The spin-torque current is trying to align the spin of all electrons of the TIA assembly in one direction over the whole sample.

For example, if the spin directions of the TIA assembly are different in two close regions in a metal, there is a mutual diffusion of electrons between these two regions. Since the spin direction of the diffused electrons is different, they cause a spin torque (See previous chapter), which rotates the spin direction of the TIA assembly. Because of this spin torque, the spin directions of each region are turning toward each other.

It should be noted that the spin-torque current is always accompanied by an additional spin relaxation. This means that the spin life time becomes shorter in regions where the spin-torque current flows.

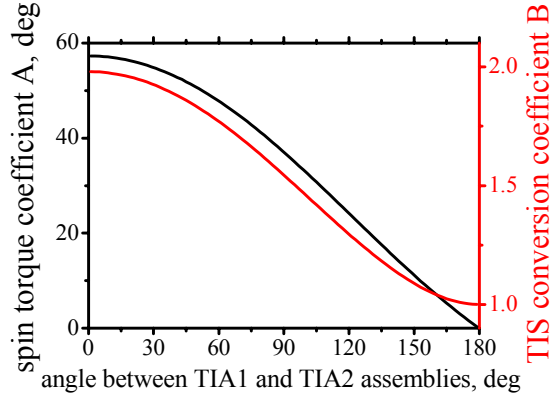

*Fig.11 Spin torque coefficient A from Eq (65) and TIS conversion coefficient B from Eq. (21) as functions of the initial angle between the TIA1 and TIA2 assemblies.*

In the case of a 1D geometry the spin torque is calculated from the random-walk model (See Appendix 3) as

$$\frac{\partial \vec{s}_{TIA}}{\partial t} = \frac{A}{n_{TIA}} 2D \frac{\partial n_{TIA}}{\partial x} \cdot \vec{s}_{TIA} \times \left[ \vec{s}_{TIA} \times \frac{\partial \vec{s}_{TIA}}{\partial x} \right] \quad (67)$$

where $\vec{s}_{TIA}$ is a unit vector directed along the spin direction of the TIA assembly, D is the diffusion coefficient, $n_{TIA}$ is the number of electrons in the TIA assembly and coefficient A equals ~57 (See Fig 11) .

The spin-torque current is accompanied by a conversion of electrons into the TIS assembly (additional spin relaxation) at the rate

$$\frac{\partial n_{TIS}}{\partial t} = B \cdot 2D \left( 0.5 \cdot n_{TIA} \cdot \vec{s}_{TIA} \cdot \frac{\partial^2 \vec{s}_{TIA}}{\partial x^2} + \frac{\partial n_{TIA}}{\partial x} \vec{s}_{TIA} \cdot \frac{\partial \vec{s}_{TIA}}{\partial x} \right) \quad (68)$$

where the coefficient B equals ~2 (See Fig 11).

In the case when the spin accumulation exponentially decays along the diffusion distance

$$n_{TIA} = n_{TIA0} e^{-\frac{x}{l_{spin}}}$$

where $l_{spin}$ is the spin diffusion length, Eqs (68) and (67) are simplified to

$$\frac{\partial \vec{s}_{TIA}}{\partial t} = -2D \cdot \frac{A}{l_{spin}} \cdot \vec{s}_{TIA} \times \left[ \vec{s}_{TIA} \times \frac{\partial \vec{s}_{TIA}}{\partial x} \right] \quad (69)$$

$$\frac{\partial n_{TIS}}{\partial t} = B \cdot 2D \cdot n_{TIA} \cdot \vec{s}_{TIA} \cdot \left( 0.5 \frac{\partial^2 \vec{s}_{TIA}}{\partial x^2} - \frac{1}{l_{spin}} \cdot \frac{\partial \vec{s}_{TIA}}{\partial x} \right) \tag{70}$$

**10. Physical origin of the spin-transfer torque.**

The spin-transfer torque is the torque acting on the magnetization of a ferromagnetic electrode of the magnetic tunnel junction (MTJ), when the electrical current flows between the electrodes of the MTJ. The spin-transfer torque may cause either a magnetization precession in the ferromagnetic electrode or a reversal of the magnetization of the electrode. The spin-transfer torque is utilized as a recording mechanism in the spin-transfer-torque magnetic random access memory (STT-MRAM). It might be possible that in the future the STT-MRAM will be a universal memory, which may replace high-density non-volatile memory (hard disks, Flash memory) and high-density fast-speed memory (DRAM, SRAM).

The MTJ is a basic cell of the STT-MRAM memory. The MTJ consists of two ferromagnetic metals and a thin isolator (a tunnel barrier) between them. The magnetization of one ferromagnetic metal is pinned by an exchange field with an antiferromagnetic layer and the magnetization of this layer can not be reversed. This ferromagnetic layer is called the "pinned" layer. The magnetization of the second ferromagnetic layer may have two stable directions along its easy axis. This ferromagnetic region is called the "free" layer. A bit of data in the MTJ cell is stored by means of two opposite magnetization directions of the "free" layer. Due to the spin transfer torque the magnetization of the "free" layer can be reversed by an electrical current. Therefore, a bit of data can be recorded.

The spin transfer torque was theoretically predicted[21,22] in 1993. The first experimental demonstration[23,24] of the spin transfer torque in MTJ was in 2004. Despite the earlier success in the development of the STT-MRAM, up to date the high-density STT-MRAM is not yet commercially available. It is still difficult to optimize the STT-MRAM for high performance and to achieve the required reliability. One reason for these difficulties is poor understanding of physical mechanisms, which govern the spin transfer torque.

In the following the physical origin of the spin transfer torque is explained using the model of TIS/TIA assemblies. It is important to emphasis that an essential feature of the spin transfer torque is the non-zero angle between the spin direction of conduction electrons of the TIA assembly and the spin direction of the local d-electrons in the ferromagnetic electrode when a current flows through the MTJ. The existence of a non-zero angle between the spin directions of the d- and conduction electrons was assumed by Huney et.al.[25] for a description of the spin transfer torque using the classical model. The proposed model gives the different description. As was explained in Chapter 6, in equilibrium in a ferromagnetic metal the conduction electrons are in both TIA and TIS assemblies. The spin direction of the TIA assembly is aligned along the spin direction of the d-electrons. In the case when the magnetization directions of the ferromagnetic electrodes of the MTJ are different, the spin directions of the TIA assemblies in each electrode are different as well. As was shown in Chapter 7, in the case when the spin direction varies over different points in space, the spin-torque current flows between these points and it tries to align all spins in the same direction. The spin-torque acts only on conduction electrons. Therefore, the spin-torque current turns the spin direction of the TIA assembly away from the spin direction of the d-electrons. The exchange interaction between the conduction electrons and the d-electrons causes a precession of spins of d-electrons and a precession of spins of electrons of the TIA assembly around a common axis, when the angle between the spin directions of the d- and conduction electrons is non-zero. In the case of a sufficiently large spin-torque current, the spin direction of the d-electrons may be reversed.

The spin-transfer torque has contributions from several other effects. For better understanding, the effects leading to the spin-transfer torque may be divided into several events or steps. Firstly, under applied voltage the drift current flows from one ferromagnetic electrode to other electrode. The spin polarization of each electrode is non-zero and the conduction electrons are in both TIS and TIA assemblies. This means that the drift current flowing in the electrodes is spin-polarized and in the drift flow there are some electrons from both TIA and TIS assemblies. The direction of spin polarization of the drift current in each electrode is different. Spins are accumulated at a tunnel barrier and a spin diffusion current flows away from the tunnel barrier inside each ferromagnetic layer. The spin diffusion current decays exponentially as it flows away from the tunnel barrier.

The spin direction of the TIA assembly is different in ferromagnetic electrodes. Also, there is a gradient of the spin accumulation. As was shown in the previous chapter, these two conditions are sufficient for a spin-torque current to flow between the electrodes. Since the spin-torque current is linearly proportional to the gradient of the spin accumulation, it is largest near the tunnel barrier and decays exponentially as it flows away from the tunnel barrier.

In the absence of current, the spin direction of the TIA assembly is along to the spin direction of the d-electrons. The spin torque current turns the spin direction of the TIA assembly away from the spin direction of the d-electrons. The exchange interaction between the d-electrons and the electrons of the TIA assembly leads to a spin precession of d-electrons and a spin precession of the electrons of TIA assembly. The damping of these precessions induces a torque acting on the d-electrons and a torque acting on the conduction electrons of the TIA assembly. These torques are directed toward each other and they are trying to align the d-electron and the conduction electrons of TIA assembly back along each other. The torque acting on d-electrons turns spins of the d-electrons away from the easy axis direction. In the case when this torque is sufficiently large, the spin direction of the d-electrons may be reversed. As was mentioned above, this effect is used for recording data in the STT-MRAM memory. In the case when the torque is not sufficient for the magnetization reversal, there is a stable precession of the d-electrons and the electrons of the TIA assembly. Since the resistance of the MTJ depends on the relative spin directions of electrons of the TIA assemblies at different sides of the tunnel barrier, the resistance of the MTJ may be modulated following the precession of electrons of the TIA assembly. The precession frequency is commonly in the microwave spectrum region and the DC current flowing through the MTJ can be modulated at a microwave frequency. This method is used to generate microwave oscillations in the microwave torque oscillator[26].

There are several different torques acting on the d-electrons and the electrons of the TIA assembly. As was mentioned above, the electrons of the TIA assembly experience the torque, which due to the spin-torque current, and the torque, which is due to exchange interaction with d-electrons. Also, the electrons of the TIA assembly experience another torque. As was shown in Chapter 4 and Chapter 6, the electrons of the TIS assembly are continuously converted into the TIA assembly, because of their exchange interaction with the d-electrons. The spin direction of the converted electrons coincides with the spin direction of the d-electrons. When there is an angle between the spin directions of the electrons of the main TIA assembly and the d-electrons, the spin direction of the converted electrons is different from the spin direction of the main TIA assembly. As was shown in Chapter 7, in this case both TIA assemblies combine and the electrons of the main TIA assembly experience the spin torque, which is trying to align the spin direction of the main TIA assembly along the spin direction of the d-electrons.

The d-electrons experience the torque, which is due to the exchange interaction with the electrons of the TIA assembly. Due to this torque the spin direction of the d-electrons may rotate away from the easy axis direction. In this case a toque is induced, which is trying to rotate the d-electrons back to be aligned along the easy axis. The exchange interaction between the d-electrons may cause another torque acting on the d-electrons. In the following its origin is explained. The spin transfer torque, the same as spin diffusion current and spin torque current, is largest near the tunnel barrier and it decreases exponentially away from the tunnel barrier. There could be a significant gradient of the spin transfer torque near the tunnel barrier. In this case the rotation angle of the different d-electrons may differ. This would cause the torque, which due to the exchange interaction between the d-electrons. The exchange interaction is trying to align the spins of all d-electrons to be parallel.

As was mentioned above, the spin transfer torque decreases exponentially with distance away of the tunnel barrier. To achieve the largest spin transfer torque and the smallest threshold current for the magnetization reversal, the thickness of the "free" layer should be as thin as possible. At least it should be thinner than the spin diffusion length in the "free" layer. As was explained in Chapter 8, the spin torque current is always accompanied by an additional spin relaxation. Therefore, the effective spin diffusion length becomes even shorter when the current flows through the MTJ.

The spin transfer torque has contributions from several physical mechanisms and effects. In order to calculate the spin transfer torque, the dynamics of the precision of the d-electrons and the electrons of the TIA assembly, the electron conversion from the TIS assembly to the TIA assembly, spin relaxation, spin and spin-torque diffusion should be calculated self-consistently. Understanding the physical mechanisms of the spin transfer torque makes it possible to optimize the STT-MRAM for better performance, which is required for its commercialization.

## 11. Conclusion.

When an electron state is filled up by two electrons of opposite spin, the spin direction of each electron is not distinguishable. Because of this fact, the spin direction of the electrons may not be conserved even in the case of spin-independent scatterings. The classical model of the spin-up/spin-down bands, which has been used for description of the magnetic properties of conduction electrons, is based on the fact that the spin direction of a conduction electron is conserved for a long time and it does not consider the fact that the spin direction may rotate after spin-independent scatterings. The proposed model of TIS/TIA assemblies used for description of the magnetic properties of conduction electrons takes into account the fact, that the spin of conduction electrons is not conserved after spin-independent scatterings. In the proposed model the electrons are divided into TIS and TIA assemblies. The TIA assembly contains only "spin" states with the same spin direction. The TIS assembly contains "full" states, "empty" states and "spin" states. In the TIS assembly "spin" states can have any spin direction with equal probability. The relative number of "full", "empty" and "spin" states is fixed by spin –independent scatterings and it depends on the electron energy.

There are significant differences between the classical and the proposed models. As was shown in Chapter 7, the basic assumption of the classical model, that electrons of only two opposite spin directions can coexist for a long time, is incorrect. This is because the spin rotation during frequent spin-independent scatterings mixes up all electrons into one TIS and one TIA assembly within a short time.

Despite all differences, the proposed model of TIS/TIA assemblies should not be considered as a contradiction of the classical model of spin-up/spin-down bands, but rather as its extension. The similarities between the models can be seen especially in the case when the direction of spin accumulation is the same throughout whole metal. For example, there are similarities in the description of Pauli paramagentism and of the properties of conduction electrons in a ferromagnetic metal in the two models. Of course, there are still some differences, because the classical model ignores the spin rotation during spin-independent scatterings. Some correlations between the models can be made. For example, the measured parameters of the classical model are spin accumulation and charge accumulation. In the proposed model the spin accumulation is due to the electrons of the TIA assembly, and the TIS assembly represents the electrons, in which there is no spin accumulation. In contrast to the classical model, where it is possible to describe spin transport and spin relaxation by a difference of Fermi energies for spin-up and spin-down electrons, the Fermi energies of assemblies are the same. Still since the spin-independent scatterings do not change the number of electrons in each assembly, it is possible to describe each assembly by an individual chemical potential.

The proposed model has an advantage in the description of the case, when the spin accumulation has different spin directions at different points in a metal. It is very difficult to describe such a case by the classical model, but the description by the proposed model is straightforward. For example, new effects such as the spin torque and the spin-torque current were predicted and the physical origin of spin transfer torque was described within the proposed model of the TIA and TIS assemblies.

The proposed model of TIA and TIS assemblies might be useful for the optimization of spintronics devices in order for them to reach the required parameters for commercialization. Especially, this theory may be advantageous for optimizing the devices, in which the magnetization or spin direction of spin accumulation changes over time or over space. The STT-MRAM is one important example of such a device.

**Appendix 1.**

In the appendix the equilibrium amount of "spin" states in the TIS and TIA assemblies will be calculated. Also, it will be proved that spin-independent scatterings do not change the number of electrons in the TIA and TIS assemblies.

Even though a spin-independent scattering may rotate the spin of an individual electron, spin-independent scatterings do not affect the total spin of all conduction electrons. This is because, all conduction electrons can be considered as a closed system. The spin-independent scatterings are events occurring inside this closed system and only a source from the outside can change its spin.

The conservation of the total spin during spin-independent scatterings also means the conservation of the projection of the total spin on any axis. The projection of the total spin on some axis can be calculated by summing up the spin projections of each electron on this axis.

For example, if $p_{spin}(\theta)$ is the probability of a "spin" state to have an angle between their spin direction and the z-axis in the interval $(\theta, \theta+d\theta)$, the probability that a "spin" state has spin direction along (spin-up) and opposite (spin-down) to the z-axis is calculated as

$$p_\uparrow = \int_0^\pi p_{spin}(\theta) \cdot \cos^2(\theta/2) d\theta$$
$$p_\downarrow = \int_0^\pi p_{spin}(\theta) \cdot \sin^2(\theta/2) d\theta \tag{A1.1}$$

Both integrals should not be changed during a spin-independent scattering, including scatterings in which spin direction is not conserved (shown in Fig.2).

In the TIS assembly there is an equal number of "spin" states with all spin directions, therefore the distribution of spin directions can be expressed as

$$p_{spin,TIS}(\theta) = 0.5 \cdot p_{TIS} \cdot \sin(\theta) \tag{A1.2}$$

where $p_{TIS} = \int_0^\pi p_{spin,TIS}(\theta) \cdot d\theta$ is the probability of "spin" states to be in the TIS assembly.

Since a "spin" state of the TIS assembly can any spin direction with equal probability, the sum of the projections of spin on any axis over all "spin" states should not depend on the axis direction. For example, the spin-up and spin-down projections (Eqs (A1.1)) should be the same.

$$p_\uparrow = p_\downarrow \tag{A1.3}$$

Substituting Eq. (A1.2) into Eq.(A1.2), this statement can be verified for the TIS assembly. Eq. (A1.4) is equivalent to

$$0 = p_\uparrow - p_\downarrow = \int_0^\pi p_{spin,TIS}(\theta) \cdot \cos^2(\theta/2) d\theta - \int_0^\pi p_{spin,TIS}(\theta) \cdot \sin^2(\theta/2) d\theta = 2\int_0^\pi p_{spin,TIS}(\theta) \cdot \cos(\theta) d\theta = 0$$
$$\tag{A1.4}$$

In the TIA assembly all spins are directed in the same direction. Therefore, the probability to find an electron of this assembly with spin directed opposite to the spin direction of the assembly is zero

$$0 = p_\downarrow = \int_0^\pi p_{spin,TIA}(\theta) \cdot \sin^2(\theta/2) d\theta \tag{A1.5}$$

In the case when the spin direction of the TIA assembly is along the z-axis ($\theta=0$), the distribution of spin directions in the TIA assembly is described as

$$p_{spin,TIA}(\theta) = p_{TIA} \cdot \delta(\theta) \tag{A1.6}$$

where $\delta(\theta)$ is the delta-function.

If at some moment in time there is an arbitrary distribution of "spin" states with some arbitrary spin directions, after a short time electrons are scattered and redistributed into one TIA assembly (Eq. (A1.6)) and one TIS assembly (Eq.A1.5). Knowing the spin distribution $p_{spin}(\theta)$ prior to equilibrium, in equilibrium the probability for a "spin" state to be in the TIA assembly can be calculated from Eqs.(A1.4) and (A1.5) as

$$p_{TIA} = p_\uparrow - p_\downarrow = 2\int_0^\pi p_{spin}(\theta) \cdot \cos(\theta) d\theta \qquad (A1.7)$$

It should be noted that according to Eq. (A1.4) the "spin" states of the TIS assembly do not contribute into the integral of Eq. (A1.7).

The probability for a "spin" state to be in the TIS assembly can be calculated from Eq. (A1.7) as

$$p_{TIS} = 1 - p_{TIA} = 1 - 2\int_0^\pi p_{spin}(\theta) \cdot \cos(\theta) d\theta \qquad (A1.8)$$

It should be noted that Eq (A1.7) is asymmetrical in respect time reversal (the integral changes its sign when time is reversed) and Eq . (A1.8) is symmetrical. The electrons of the TIS assembly do not contribute to the integral (A1.7) and the electrons of the TIA assembly do not contribute to the integral (A1.8). This means that spin-independent scatterings do not change the probability for an electron to be in one of assemblies. An equivalent statement is that the spin-independent scatterings do not change the number of electrons in each assembly.

**Appendix 2**

In this appendix the solution of the Landau-Lifshiz equation (Eq.(34)) is obtained in a form, which is convenient to use for the description of the conversion of "spin" states from the TIS assembly to TIA assembly in the presence of a magnetic field (See chapter 5).

The Landau-Lifshiz equation for the spin vector of a conduction electron can be obtained from Eq.(34) as

$$\frac{d\vec{S}}{dt} = -\gamma \vec{S} \times \vec{H} - \lambda \cdot g \cdot \mu_B \vec{S} \times (\vec{S} \times \vec{H}) \qquad (A2.1)$$

We assume that the magnetic field is applied along the z-axis. Introducing new unknowns

$$S_+ = S_x + i \cdot S_y \quad S_- = S_x - i \cdot S_y \qquad (A2.2)$$

into Eq. (A2.1) gives

$$\frac{dS_+}{dt} = i\omega_L S_+ - \frac{2}{t_\lambda} S_z S_+$$

$$\frac{dS_-}{dt} = -i\omega_L S_- - \frac{2}{t_\lambda} S_z S_- \qquad (A2.3)$$

$$\frac{dS_z}{dt} = \frac{2}{t_\lambda} S_+ S_-$$

where $\omega_L = \gamma H_z$ is the Larmor frequency. A solution of Eqs. (A2.3) can be found in the form

$$\begin{pmatrix} S_+ \\ S_- \end{pmatrix} = S_{xy} \begin{pmatrix} e^{i\omega_L t} \\ e^{-i\omega_L t} \end{pmatrix} \quad (A2.4)$$

where Sxy is the projection of the spin vector onto the xy-plane. The solution (A2.4) describes the precession of spin in the xy-plane with the Larmor frequency. The Larmor frequency does not depend on the angle between the magnetic field and the spin vector. Substituting Eqs.(A2.4) into Eqs.(A2.3) gives

$$\frac{dS_{xy}}{dt} = -\frac{2}{t_\lambda} S_z S_{xy}$$
$$\frac{dS_z}{dt} = \frac{2}{t_\lambda} H_z S_{xy}^2 \quad (A2.5)$$

where $t_\lambda$ is the precession damping time (Eq.(37)). The solution of Eqs. (A2.5) is

$$\begin{pmatrix} S_z \\ S_{xy} \end{pmatrix} = \frac{1}{2}\begin{pmatrix} \cos(\theta(t)) \\ \sin(\theta(t)) \end{pmatrix} \quad (A2.6)$$

where θ(t) is the angle between the magnetic field and the direction of spin. θ(t) decreases with time, because of the precession damping. Substituting Eq. (A2.6) into Eqs. (A2.5) gives the angular speed of the decrease of the angle between the magnetic field and the spin vector as

$$\frac{d\theta(t)}{dt} = -\frac{1}{t_\lambda} \cdot \sin(\theta(t)) \quad (A2.7)$$

the solution of Eq. (A2.7) is

$$\tan\left(\frac{\theta}{2}\right) = e^{-\frac{t}{t_\lambda} + const} \quad (A2.8)$$

The Eq. (A2.8) describes the rotation of the spin towards the direction of the magnetic field due to the precession damping

## Appendix 3.

In this appendix the spin torque induced by the spin-torque current is calculated in a 1D geometry based on the random-walk model. The purpose of the following calculations is to study the basic properties of the spin torque current.

Using the random-walk model it is possible to calculate the number of electrons diffusing from one point to another. Using this number, the spin torque is calculated using Eq.(65)

Let us consider the spin diffusion between 3 points $x, x-\Delta x, x+\Delta x$ where $\Delta x \to 0$. Each point has a different spin direction and a different amount of electrons in the TIA assembly. As a result of the random walk, the number of particles, which diffuse from a point in some direction (for example, from point x towards point x+Δx), per unit time is proportional to the number of particles at that point

$$\left(\frac{\partial n}{\partial t}\right)_{x \to x+\Delta x} = D\frac{n(x)}{\Delta x^2} \quad (A3.1)$$

where D is the diffusion constant and n(x) is the density of particles at point x.

In the case when spin relaxation is weak and the conversion of electrons from the TIA to the TIS assembly is slow, it is possible to consider the diffusion of electrons of the TIA and the TIS assemblies as independent. Than, the number of electrons of the TIA assembly, that diffuse from point x+Δx to point x, is calculated as

$$\left(\frac{\partial n_{TIA}}{\partial t}\right)_{x+\Delta x \to x} = D \frac{n_{TIA}(x+\Delta x)}{\Delta x^2} \qquad (A3.2)$$

The spin direction and the number of electrons in the TIA assembly at point x+Δx can be approximated as

$$\vec{s}_{TIA}(x+\Delta x) = \vec{s}_{TIA}(x) + \frac{\partial \vec{s}_{TIA}}{\partial x}\Delta x + 0.5\frac{\partial^2 \vec{s}_{TIA}}{\partial x^2}\Delta x^2$$

$$n_{TIA}(x+\Delta x) = n_{TIA}(x) + \frac{\partial n_{TIA}}{\partial x}\Delta x \qquad (A3.3)$$

Since the spin directions at points x and x+Δx are different, electrons, that diffuse from point x+Δx, cause a spin torque on the electrons at point x. This spin torque is calculated by substitution Eq (A3.2),(A3.3) into Eq.(65) as

$$\left(\frac{\partial \vec{s}_{TIA}}{\partial t}\right)_{x+\Delta x \to x} = \frac{A}{n_{TIA}} D \frac{\left(n_{TIA} + \frac{\partial n_{TIA}}{\partial x}\Delta x\right)}{\Delta x^2} \cdot \vec{s}_{TIA} \times \left[\vec{s}_{TIA} \times \frac{\partial \vec{s}_{TIA}}{\partial x}\Delta x\right] \qquad (A3.4)$$

The spin torque is accompanied by a conversion of electrons from the TIA into the TIS assembly. The conversion rate is calculated by substitution Eq (A3.2),(A3.3) into Eq.(66) as

$$\left(\frac{\partial n_{TIS}}{\partial t}\right)_{x+\Delta x \to x} = B \cdot D \frac{\left(n_{TIA} + \frac{\partial n_{TIA}}{\partial x}\Delta x\right)}{\Delta x^2} \left(\vec{s}_{TIA} \cdot \frac{\partial \vec{s}_{TIA}}{\partial x}\Delta x + 0.5 \cdot \vec{s}_{TIA} \cdot \frac{\partial^2 \vec{s}_{TIA}}{\partial x^2}\Delta x^2\right) \qquad (A3.5)$$

Electrons also diffuse from point x-Δx towards point x. Similarly, the spin torque due to this diffusion can be calculated as

$$\left(\frac{\partial \vec{s}_{TIA}}{\partial t}\right)_{x-\Delta x \to x} = -\frac{A}{n_{TIA}} D \frac{\left(n_{TIA} - \frac{\partial n_{TIA}}{\partial x}\Delta x\right)}{\Delta x^2} \cdot \vec{s}_{TIA} \times \left[\vec{s}_{TIA} \times \frac{\partial \vec{s}_{TIA}}{\partial x}\Delta x\right] \qquad (A3.6)$$

and the electron conversion into the TIS assembly can be calculated as

$$\left(\frac{\partial n_{TIS}}{\partial t}\right)_{x-\Delta x \to x} = B \cdot D \frac{\left(n_{TIA} - \frac{\partial n_{TIA}}{\partial x}\Delta x\right)}{\Delta x^2} \left(-\vec{s}_{TIA} \cdot \frac{\partial \vec{s}_{TIA}}{\partial x}\Delta x + 0.5 \cdot \vec{s}_{TIA} \cdot \frac{\partial^2 \vec{s}_{TIA}}{\partial x^2}\Delta x^2\right) \qquad (A3.7)$$

Summing up Eqs. (A3.4) and (A3.6), the spin torque, which the electrons of the TIA assembly experience due to the diffusion from the surroundings, can be calculated as

$$\frac{\partial \vec{s}_{TIA}}{\partial t} = \left(\frac{\partial \vec{s}_{TIA}}{\partial t}\right)_{x+\Delta x \to x} + \left(\frac{\partial \vec{s}_{TIA}}{\partial t}\right)_{x-\Delta x \to x} = \frac{A}{n_{TIA}} 2D \frac{\partial n_{TIA}}{\partial x} \cdot \vec{s}_{TIA} \times \left[\vec{s}_{TIA} \times \frac{\partial \vec{s}_{TIA}}{\partial x}\right] \quad (A3.8)$$

Summing up Eqs. (A3.5) and (A3.7), the conversion rate from the TIA to the TIS assembly due to spin-torque current can be calculated as

$$\frac{\partial n_{TIS}}{\partial t} = \left(\frac{\partial n_{TIS}}{\partial t}\right)_{x+\Delta x \to x} + \left(\frac{\partial n_{TIS}}{\partial t}\right)_{x-\Delta x \to x} = B \cdot 2D \left(0.5 \cdot n_{TIA} \cdot \vec{s}_{TIA} \cdot \frac{\partial^2 \vec{s}_{TIA}}{\partial x^2} + \frac{\partial n_{TIA}}{\partial x} \vec{s}_{TIA} \cdot \frac{\partial \vec{s}_{TIA}}{\partial x}\right)$$
(A3.9)